\documentclass[]{spie}  %>>> use for US letter paper
\usepackage{amsmath,amsfonts,amssymb, bm}
\usepackage{graphicx}
\usepackage[position = top]{subfig}
\usepackage[colorlinks=true, allcolors=blue]{hyperref}
\title{Low Intensity LiDAR using Compressed\\ Sensing and a Photon Number Resolving Detector}

\author[a,b]{Yoni Sher}
\author[b]{Lior Cohen, Daniel Istrati, Hagai S. Eisenberg}
\affil[a]{School of Computer Science and Engineering, Hebrew University of Jerusalem, Jerusalem 91904, Israel}
\affil[b]{The Racah Institute of Physics, Hebrew University of Jerusalem, Jerusalem 91904, Israel}

\authorinfo{Further author information: (Send correspondence to H.S.E.)\\H.S.E: E-mail: hagai.eisenberg@mail.huji.ac.il\\  Y.S.: E-mail: yoni.sher@mail.huji.ac.il}

\begin{document} 
\maketitle

\begin{abstract}
LiDAR (laser based radar) systems are a major part of many new real-world interactive systems, one of the most notable being autonomous cars. The current market LiDAR systems are limited by detector sensitivity: when output power is at eye-safe levels, the range is limited. Long range operation also slows image acquisition as flight-time increases.  
We present an approach that combines a high sensitivity photon number resolving diode with machine learning and a micro-mechanical digital mirror device to achieve safe and fast long range 3D scanning. 
\end{abstract}

\keywords{LiDAR, compressed sensing, photon counting, imaging}

\section{INTRODUCTION}
\label{sec:intro}  % \label{} allows reference to this section

\textbf{Li}ght \textbf{D}etection \textbf{A}nd \textbf{R}anging (LiDAR) (aka laser based radar) systems are a rising sensory capability with applications to many real-world interactive systems. From 3D mapping for augmented reality, urban planning, agriculture, autonomous navigation and autonomous cars to high resolution maps for research in fields from geology to archeology \cite{nex2014uav}, accurate computerized depth maps of our surroundings are becoming a sought commodity. \\
To fix the range to an object, LiDAR systems can use time-of-flight information. Current market sensors providing fast responses to optical signals are large compared to the CCD elements used in digital cameras, making arrays of such sensors  large and expensive. As a result the usual approach uses a single detector to scan the surroundings, giving one distance measurement for each angular coordinates.\\ 
This approach (single-pixel scanning) results in acquisition times that are at least linear in the number of data points acquired. While this appears to be a reasonable scaling, maintaining spatial resolution at greater distances requires increasing the angular resolution by a fixed ratio, leading to acquisition times that increase polynomially in target distance and image resolution. Since applications measure increased resolution by the number of horizontal or vertical pixels measured, the total number of data points is squared with each increase in resolution.\\
Safety and energy efficiency are another two considerations that limits single-pixel scanning LiDAR. To overcome measurement noise the output power of the laser must be increased. Passerby safety is often improved by using less harmful wavelengths, which could require increased energy output to overcome lower sensor efficiency at the chosen wavelength. Eventually safety will win out in any wavelength, and the maximum safe output intensity will affect the effective measurement range of the LiDAR. \\
We address these issues by increasing the sensitivity in two realms, using two emerging technologies: A photon number resolving detector increases our signal sensitivity without increasing output intensity, and compressed sensing allows the capture of more information with each measurement. \\
Use of a photon number resolving detector (details in section \ref{PCD}) allows the reduction of optical illumination power by several orders of magnitude. To distinguish a signal from background light we need only several photons to return from the target. A narrow wavelength filter sufficiently reduces the ambient signal to a point where it is possible to operate in broad daylight. Additionally, high sensitivity allows capturing of targets with a wide dynamic range of intensities by leveraging the temporal separation of signals returning from different distances. \\
Compressed sensing (explained in section \ref{CS}) is a machine-learning technique based on the observation that many signals contain much less information than would be suggested by the Nyquist limit. This is especially evident for naturally occurring pictures, which can easily be compressed to ratios as low as $ 95\% $ with minimal loss of accuracy; represented in the correct bases, natural images are very sparse. We could therefore expect to recover all the information an image carries with a number of measurements closer to the information content of the image, rather than the Nyquist limit. \\
Compressed sensing uses carefully chosen measurement bases to sample the signal (in this case, image) as a whole, and can then reconstruct the signal using non-linear optimization to recover the image with $ 100\% $ fidelity from a number of measurements that is logarithmic in the signal frequency. For images, the Nyquist limit is proportional to the number of pixels, so that a mega-pixel image can be reconstructed from as few as 6000 compressed sensing measurements. 

\section{PHOTON NUMBER RESOLVING DETECTORS AND SYSTEM OVERVIEW}\label{PCD}
Photon number resolving detectors (PNRDs) are single photon detectors capable of distinguishing the number of photons arriving at the detector in a short time period \cite{dolgoshein2006status}. As PNRDs we use a silicon photo-multiplier (SiPM), which is an avalanche photo-diode (APD) array with a common charge collector operating in Geiger mode. 
\subsection{Silicon photomultiplier}
The detector is an array of APD elements, each able to detect a single photon and output a constant electric charge in a short time. All the elements are connected in parallel to the same output, such that the total current from all elements corresponds to the number of photons detected by the array. \\
APDs can be used in linear mode or Geiger mode. In linear mode, the diode is placed under reverse bias below the breakdown voltage. Photons reaching the diode release electron-hole pairs that are accelerated by the electric field, during which they can hit other electrons and release more pairs. In linear mode the energy obtained from the electric current is below that required to create another pair, and the reaction decays with time. In Geiger mode, the diode is places under reverse bias voltage beyond the breakdown voltage. A photon arriving at the diode sets off a chain reaction of released electrons and holes, creating an avalanche and a larger current, increasing the detector's gain. \\

\subsection{System overview}
\begin{figure} [ht]
	\begin{center}
		\subfloat[]{\includegraphics[height=5cm]{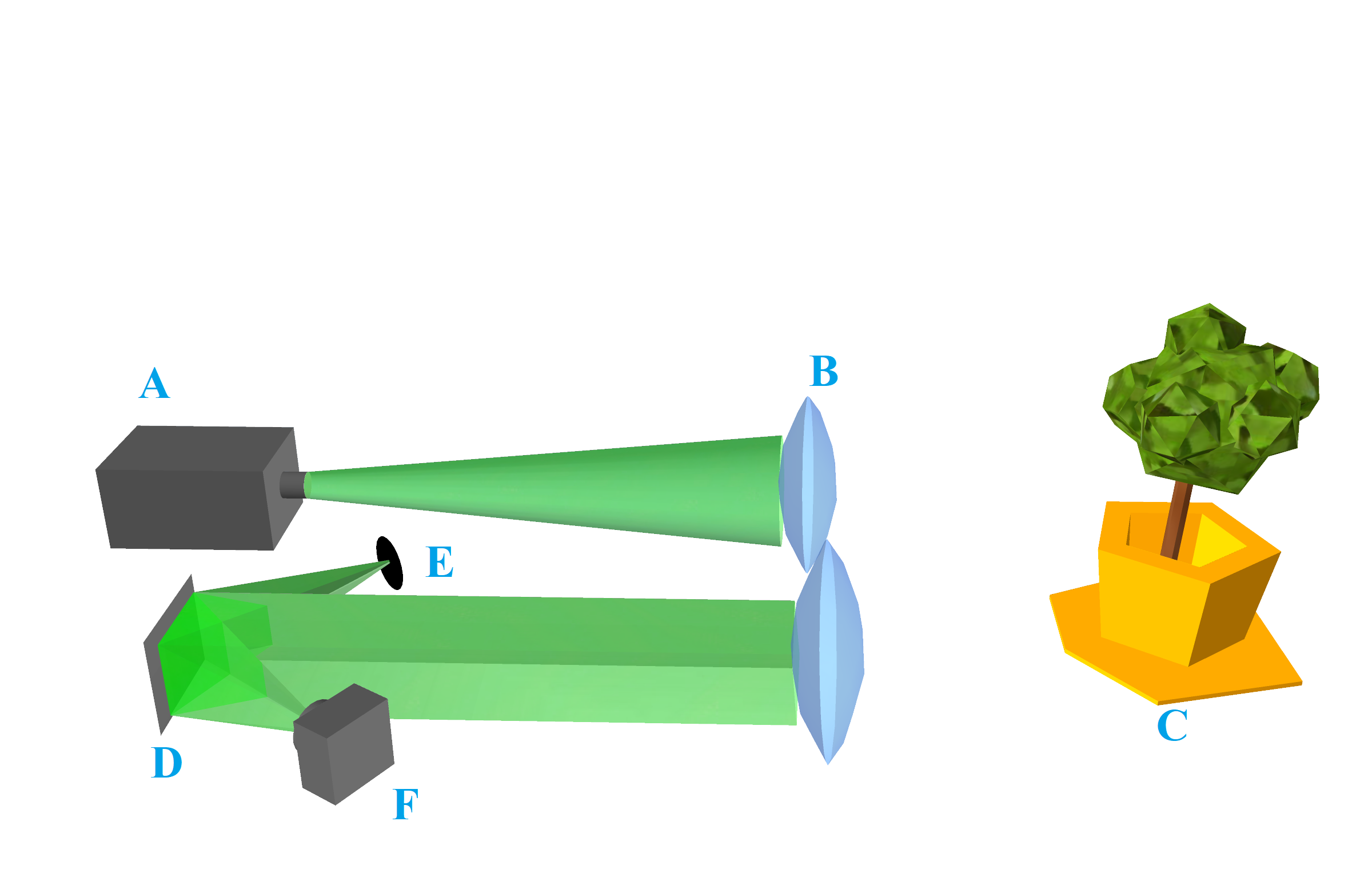}
			\label{fig_sys_ovr}}
		\hfil
		\subfloat[]{\includegraphics[height=7cm]{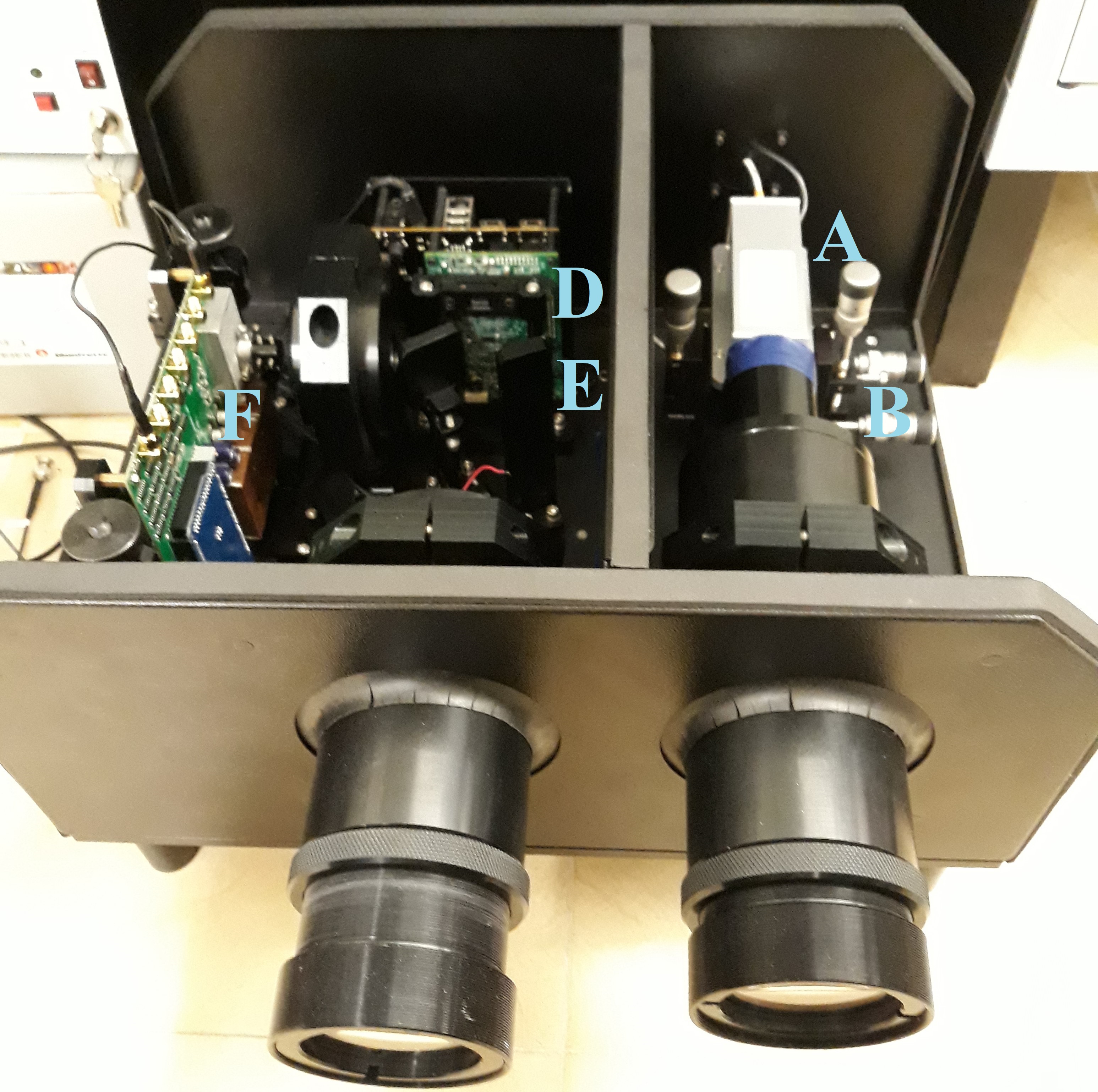}
			\label{fig_picture}}
	\end{center}
	\caption{(\textbf{a}) Illustration of the LiDAR system. (\textbf{b})Picture of the LiDAR system. \\(A) Laser (B) Holographic diffuser (C) Object (D) DMD (E) Beam-dump (F) Detector}
\end{figure}

The presented LiDAR setup is a two channel system - one for illumination and one for sensing. The illumination sourse is a Standa micro-chip Nd-Yag passively mode-locked laser frequency doubled to 532-nm wavelength with 0.5ns pulses at 1KHz repetition rate. A pulse length of 0.7nm gives a theoretical lower bound on the depth resolution of the LiDAR of about 10-cm. The beam is passed through a holographic diffuser to create a square illumination in the far field, with 30 milliradian angular spread. An optional polarizer (not shown) allows further beam attenuation if necessary. The sensing channel uses a telescope to image the returned light on a Texas Instruments DLP4500 \cite{instruments_2017} digital micro-mirror device (DMD), with a resolution of 1152 by 912 pixels. Lower resolution images can be captured by binning mirror elements into larger effective pixels. The DMD is programed to show binary masks as the chosen basis for compressed sensing (see section \ref{CS}). The light hitting the DMD is directed either to the PNRD through a beam homogenizer, or to a beam-dump coated with highly absorbent material (See figure \ref{fig_sys_ovr}, \ref{fig_picture}). \\

The DMD allows us to choose non-trivial masks and thereby use compressed sensing. The homogenizer maps every pixel of the DMD to the entire area of the detector, limiting the finite-area effects that cause non-linearity at higher intensities \cite{dolgoshein2006status} and allowing a wider dynamic range. The beam-dump  absorbs all the light returned from the target that is not part of the current measurement. This light would otherwise become particularly disruptive noise: because it is correlated with the signal we are trying to measure, it can not be distinguished from the signal and so can not be removed with standard techniques. This setup is inspired by the single photon LiDAR described in Ref. ~\citenum{howland2014compressive}.

\subsection{Data acquisition}
The signal from the detector is recorded on a fast computer controlled oscilloscope for analysis. The effective range of the LiDAR depends on the illumination power, target albedo and detector sensitivity, and this range in turn dictates the appropriate trace length we need to record after each laser pulse. The distance to the target is calculated by relating the return trip time to the speed of light, so the depth resolution depends on the temporal resolution of the oscilloscope. An example trace is presented in Fig. \ref{dTrace}. Once the data is recorded and the targets in it identified, it is split into depth frames. Each frame is taken to be a 2D image, and compressed sensing (outlined in section \ref{CS}) is used to reconstruct the respective image from the measured signals. 

\section{Compressed Sensing}\label{CS}
Compressed Sensing is a method of reconstructing signals from fewer measurements than the Nyquist limit for the highest frequency in the signal. For monochromatic images, this means reconstructing the image from fewer measurements than the number of pixels in the image. It is based on two assumptions: The image is sparse in some basis (a compressed information representation), and that we can measure in a basis incoherent with the information representation \cite{candes2007sparsity}.  \\
Sparsity gives rise to the following argument: the Shannon entropy of an image with $ n $ pixels but only $ k \ll n $ non-zero elements is \cite{howland2014compressive} 
\begin{equation}\label{entropy}
n\left [-\frac{k}{n}  \log{\frac{k}{n}} -\left(1-\frac{k}{n}\right)  \log\left(1-\frac{k}{n}\right)\right ] \approx  k \log \frac{n}{k} 
\end{equation}
We use this as a lower bound for the number of measurements required to reconstruct a signal. We will empirically show that we can get very close to this bound. \\
\subsubsection{Incoherence} \cite{candes2008introduction,howland2014compressive} Two bases are mutually incoherent if the dot product of any vector from one with any vector from the other is not too large: If $\{\phi_i\},\{\psi_j\} $  are basis of dimension $n$, the mutual incoherence is:
\begin{equation}\label{mutual_incoherence}
\mu({\phi_i},{\psi_j}) = \sqrt{n} \max_{i,j} \|\langle\phi_i, \psi_j \rangle\|
\end{equation}
This can be intuitively understood as follows: measuring in a coherent base, we only get information if we measure a quantity present in our signal. In an incoherent base, any measurement gives us information about the entire signal. \\
\subsubsection{Sparsity} Natural images tend to be sparse in frequency bases such as Fourier or wavelet. Random measurement bases will be $ C \cdot \sqrt{2 \log n} $ incoherent with any base with high probability for some $ C \approx 1 $, and are therefore suitable for compressed sensing.
\subsection{Reconstruction theory}
Let $ x $ be some signal, and $\{\phi_i\} $ be some basis in which $ x $ is $ k $-sparse. Let $ \{y_k\} $ be a set of $ m $ measurements of $ x $ in some basis $ \bm{\psi} $, and let 
\begin{equation}\label{min_measurements}
m>\mu(\bm{\phi,\psi})^2 \cdot k\cdot \log\frac{n}{\delta}
\end{equation}
Then with probability greater than $ 1-\delta $:
\begin{equation}\label{reconstruction}
x=\min_{x' \in \mathbb{R}^n} \| x' \|_1 \quad \text{s.t.} \quad y_k=\langle\phi_k,x'\rangle
\end{equation}
This implies that if we take $ m \ge k \log^2 n $ measurements of $ x $ in some random basis, we can expect to reconstruct $ x $ with high probability\cite{candes2008introduction}. \\ 

\subsection{Reconstruction practice}
\begin{figure}[ht]
	\begin{center}
		\subfloat[]{\includegraphics[height=5cm]{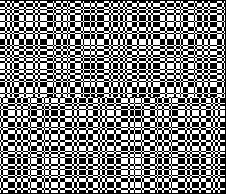}
			\label{dragonLet}}
		\hfil
		\subfloat[]{\includegraphics[height=5cm]{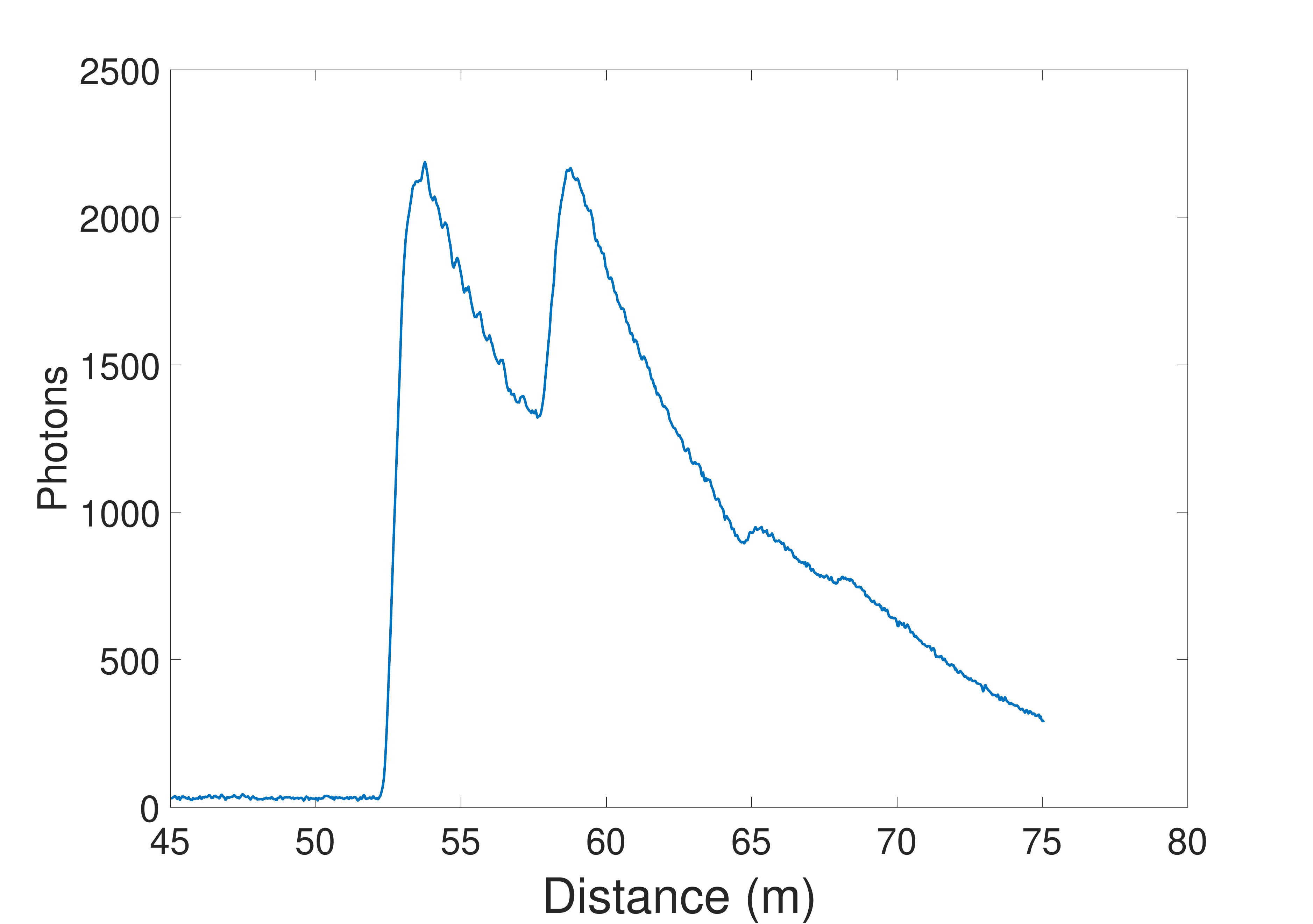}
			\label{dTrace}}	
	\end{center}
	\caption{(\textbf{a}) Example of a dragon wavelet. (\textbf{b}) Raw data from PNR detector.}
\end{figure}

Practically, $ m \approx k \log n$ measurements are usually sufficient to reconstruct the signal exactly, and signals can be approximately reconstructed from as few as $ m=O(k) $ measurements. \\
The measurements used in this work are random elements of the Dragon Wavelet group (Fig. \ref{dragonLet}), described in Ref. ~\citenum{feldman2016power}, which can be computed efficiently in $ O(n \log n) $ time and in-place for optimal memory use. The Dragon wavelet group resemble fractal noise patterns, and so are highly incoherent with natural images as well as other signals we would be likely to measure. \\
There are many algorithms available for compressed sensing reconstruction, from standard of-the-shelf algorithms (such as linear programming) implemented by every scientific computation package to optimized special-purpose methods for sparse signal reconstruction. We used the MATLAB implementation of Nestorov's algorithm (NESTA)  \cite{nesterov1983method} with total variation (TV) minimization instead of $ L_1 $ sparsity primarily for the increased robustness to noise presented by this approach \footnote{A more rigorous analysis is forthcoming}. \\
When reconstructing a 3D image, we treat each depth layer as a 2D image and then stack the layers, which allows a much simplified temporal acquisition scheme. The linear temporal scan also allows a simpler computational pipeline for reconstruction than a fully 3D compressed sensing approach would require, but does not allow the exploitation of compressed sensing for the depth dimension. Figure \ref{dTrace} presents a trace for a single mask recorded by the system, showing the number of photons received as a function of the distance. Each trace is correlated with the detector's response curve and the time of arrival and photon count are calculated.  This is one of the traces used to create Fig. \ref{CSChimny}. 

\section{Results}
We aimed the system at a variety of targets available in out laboratory's surroundings, including chimneys, signs and environmental sculptures. The amount of signal required is determined by background light: there can be anywhere from 2 photons on a bright night to 30 solar photons by day per nanosecond in the 532$ \pm $ 1nm wavelength window of the filter attached to the detector. Thus, we need to detect at least 900 photons per measurement in order to reach shot noise levels comparable to the background radiation. The current limit on the system's resolution is 64x64 pixels and stems from the accuracy of the data acquisition and recording equipment compatible with our system. Several software and hardware solutions are available to increase the resolution to the full capacity of the DMD, but have not yet been fully integrated. 

\subsection{Close targets}

\begin{figure}[ht]
	\begin{center}
		\subfloat[]{\includegraphics[height=5cm]{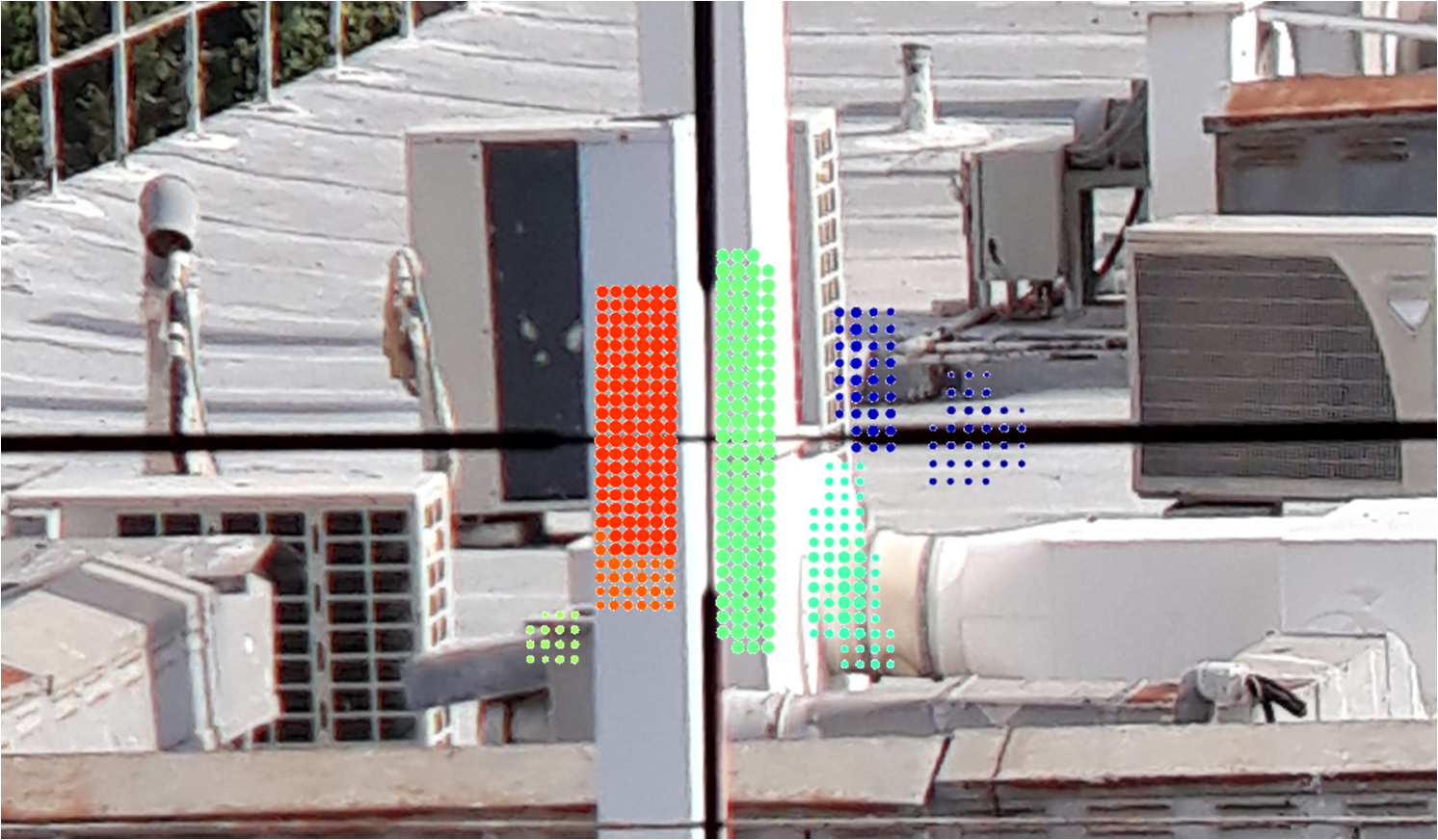}
			\label{chimFront}}
		\hfil
		\subfloat[]{\includegraphics[width=6cm]{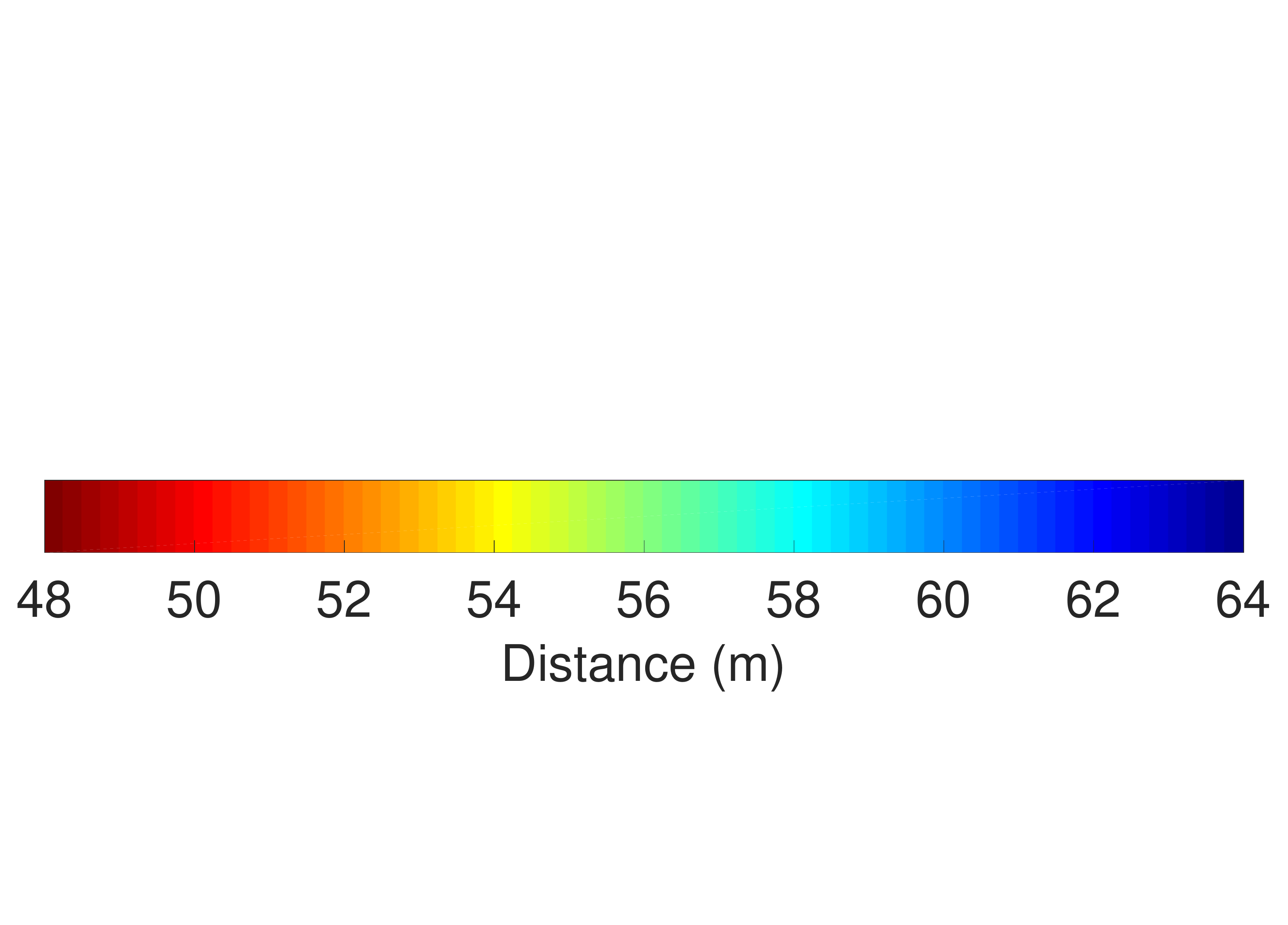}}
		\hfil
		\subfloat[]{\includegraphics[height=5cm]{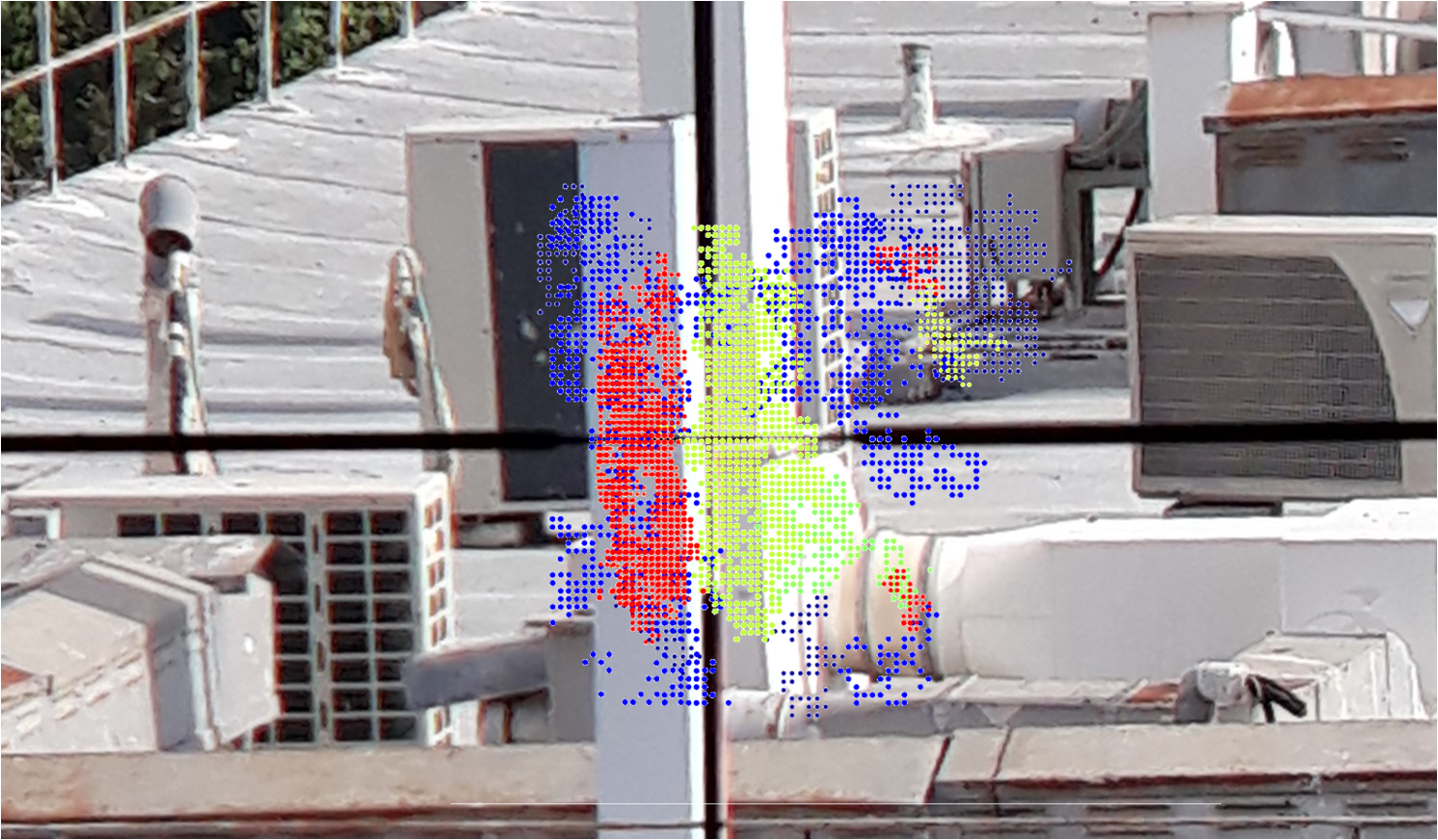}
			\label{chimFrontCS}}
		\hfil
		\subfloat[]{\includegraphics[height=5cm]{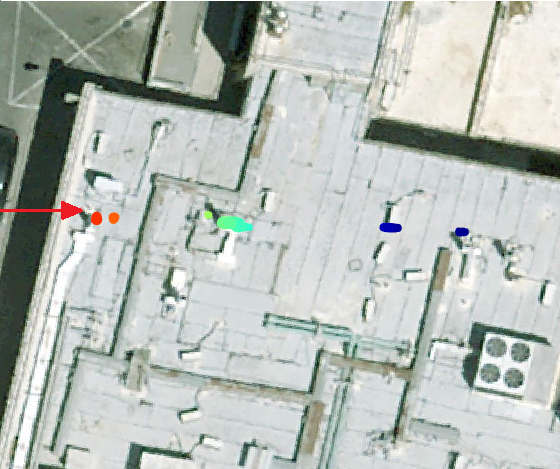}
			\label{chimTop}}
	\end{center}
	\caption{(\textbf{a}) Picture taken through LiDAR sight with raster scan overlay.  (\textbf{b}) Depth scale for LiDAR images.(\textbf{c})  Picture taken through LiDAR sight with compressed sensing overlay.  (\textbf{d}) Overhead image with overlay. The red line indicates the direction of view.}
\end{figure}

\begin{figure}[ht]
	\begin{center}
		\subfloat[]{\includegraphics[height=3.7cm]{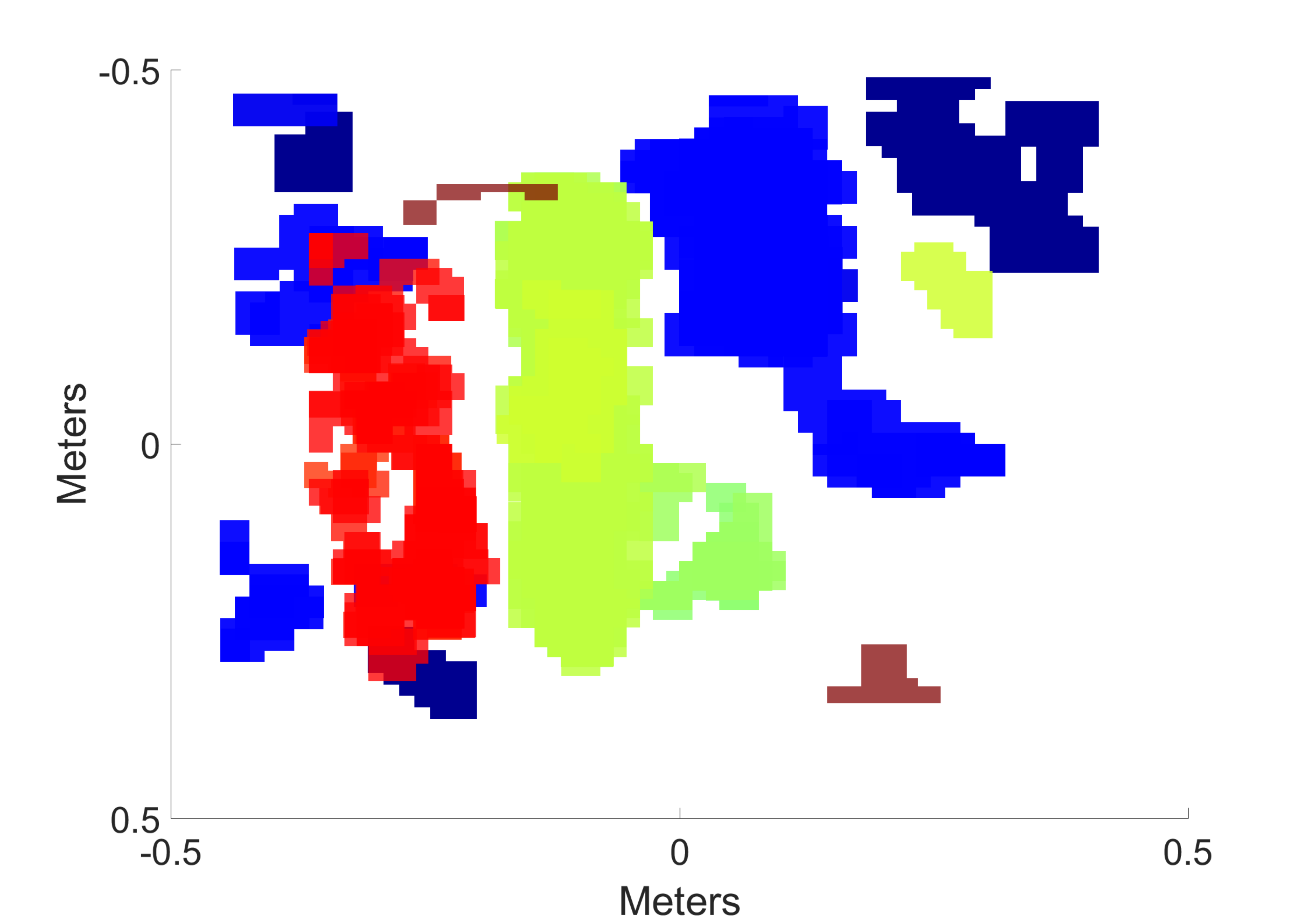}
			\label{chim_512m}}
		\subfloat[]{\includegraphics[height=3.7cm]{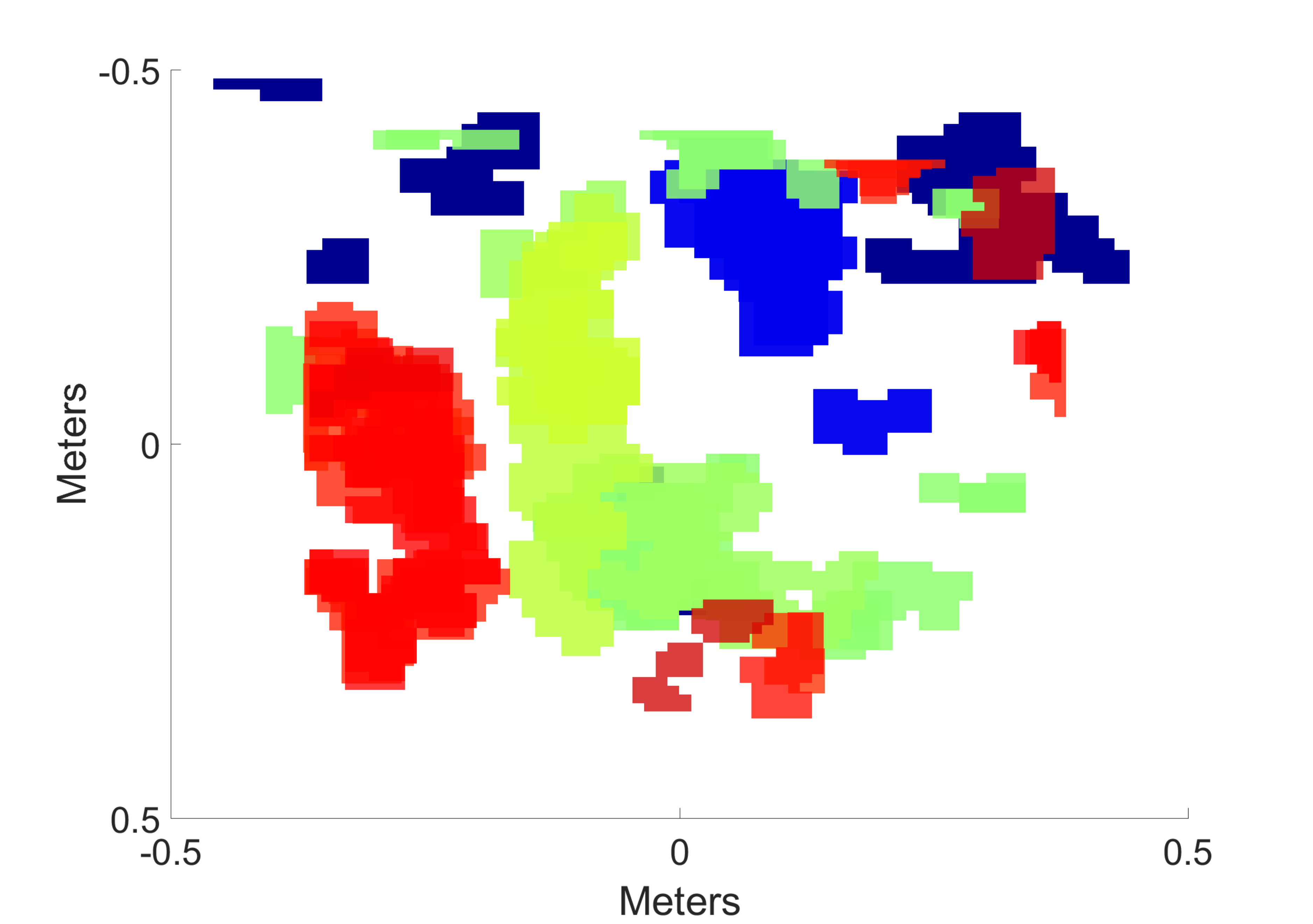}
			\label{chim_256m}}
		\subfloat[]{\includegraphics[height=3.7cm]{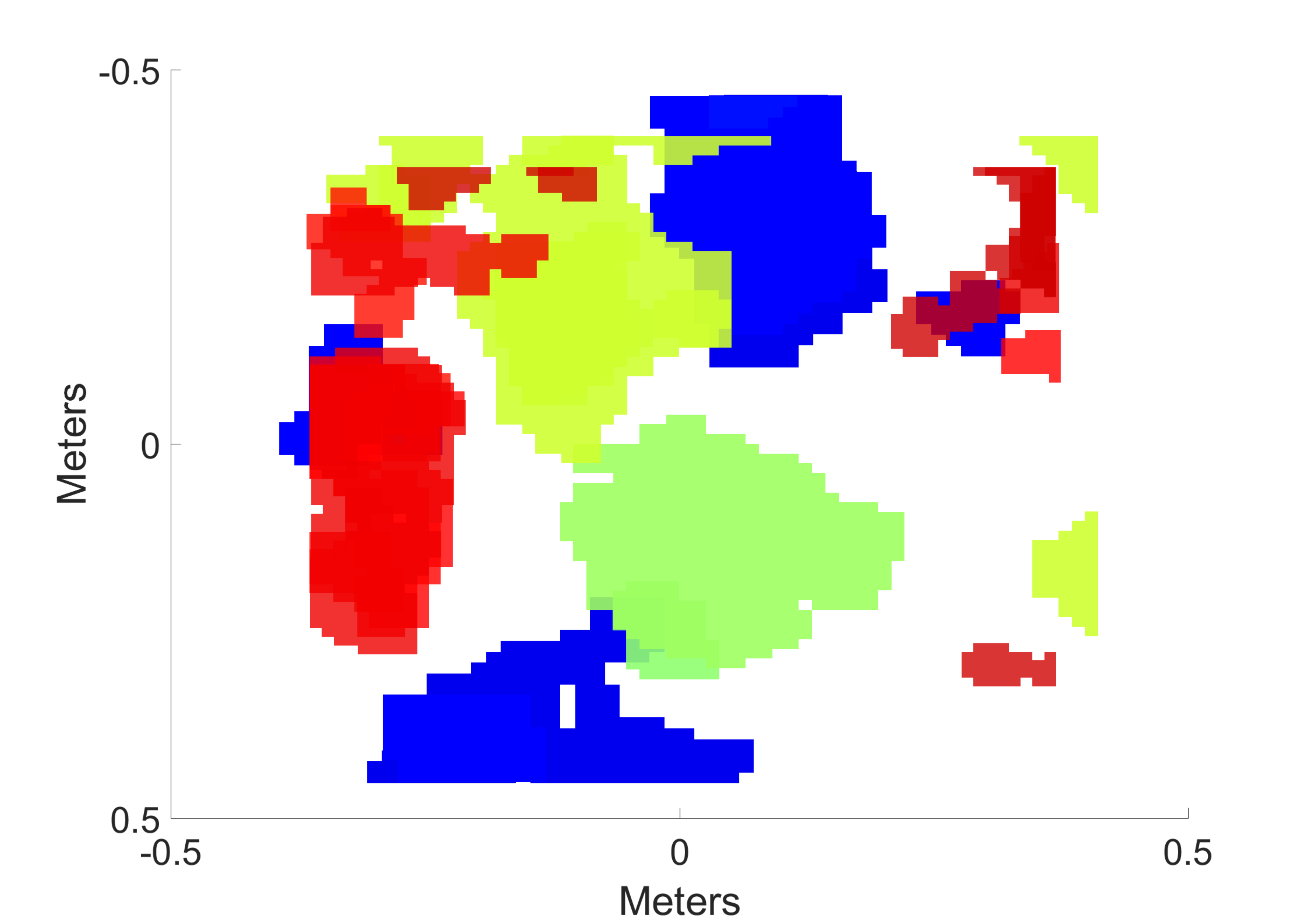}
			\label{chim_64m}}
		\subfloat[]{\includegraphics[height=3.7cm]{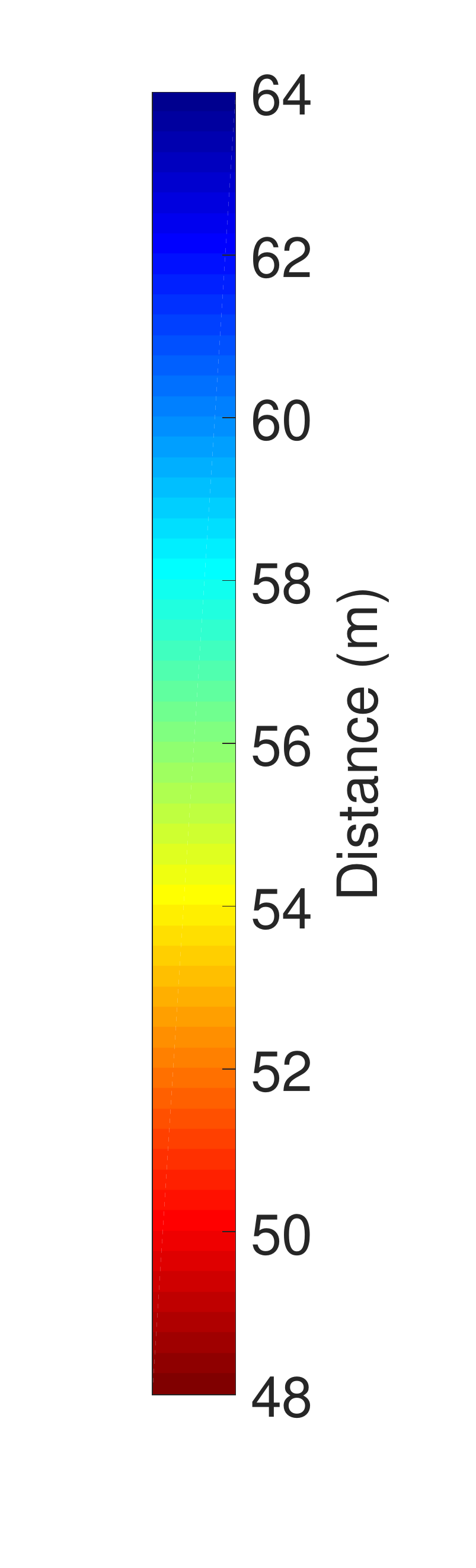}}
	\end{center}
	\caption{Comparison of reconstruction by mask number. (\textbf{a}) Reconstuction using 512 masks, 12.5\% the number of pixels. (\textbf{b}) Reconstuction using 256 masks (6.25\%). (\textbf{c}) Reconstuction using 64 masks (1.56\%). }
	\label{CSChimny}
\end{figure}

We started with a close by target in the 50-60 meter range. Figures \ref{chimFront}, \ref{chimFrontCS} show an optical camera image of several chimney pipes found across the street from our lab, and a top view of the same targets (Fig. \ref{chimTop}). The 3D depth information created by the system is superimposed on the images. Figure \ref{chimFront} is a raster scan, obtained by turning on one pixel at a time in order, with the output intensity set so that approximately 50 photons are returned from the target per pixel above the background radiation. Each pixel is measured 10 times, giving around 500 photons per pixel. This is sufficient to distinguish a target from empty space but does not reveal any more detailed information. \\
For compressed sensing, the information carried by each mask is in the mask's intensity compared to the other masks. The variation in intensity between masks is on the order of the square root of the signal frequency \cite{howland2014compressive}, which for square images is proportional to the number of pixels on each side. Thus we set the output intensity so that approximately 2,500 photons are returned per mask, with each mask covering half the image's pixels. Repeating each measurement 10 times yields 25,000 photons per mask, which gives sufficient measurement accuracy for image reconstruction. Compressed sensing allows us to vary the number of masks used to reconstruct the image. Figure \ref{CSChimny} shows a compressed sensing scan of the same image reconstructed with different numbers of masks, from 512 masks (in figure \ref{chim_512m}) down to just 64 masks (\ref{chim_64m}). Using 512 masks, the detector recorded 3,125 photons for each of the 4096 pixels, giving a theoretical SNR of approximately 3. This is far from ideal conditions, and causes incomplete and noisy reconstruction. On the other hand, the larger collection area of each compressed sensing mask allows us to image depth planes where the return signal is below one photon per pixel. \\

\subsection{Distant targets}

\begin{figure}[ht]
	\begin{center}
		\subfloat[]{\includegraphics[height=6cm]{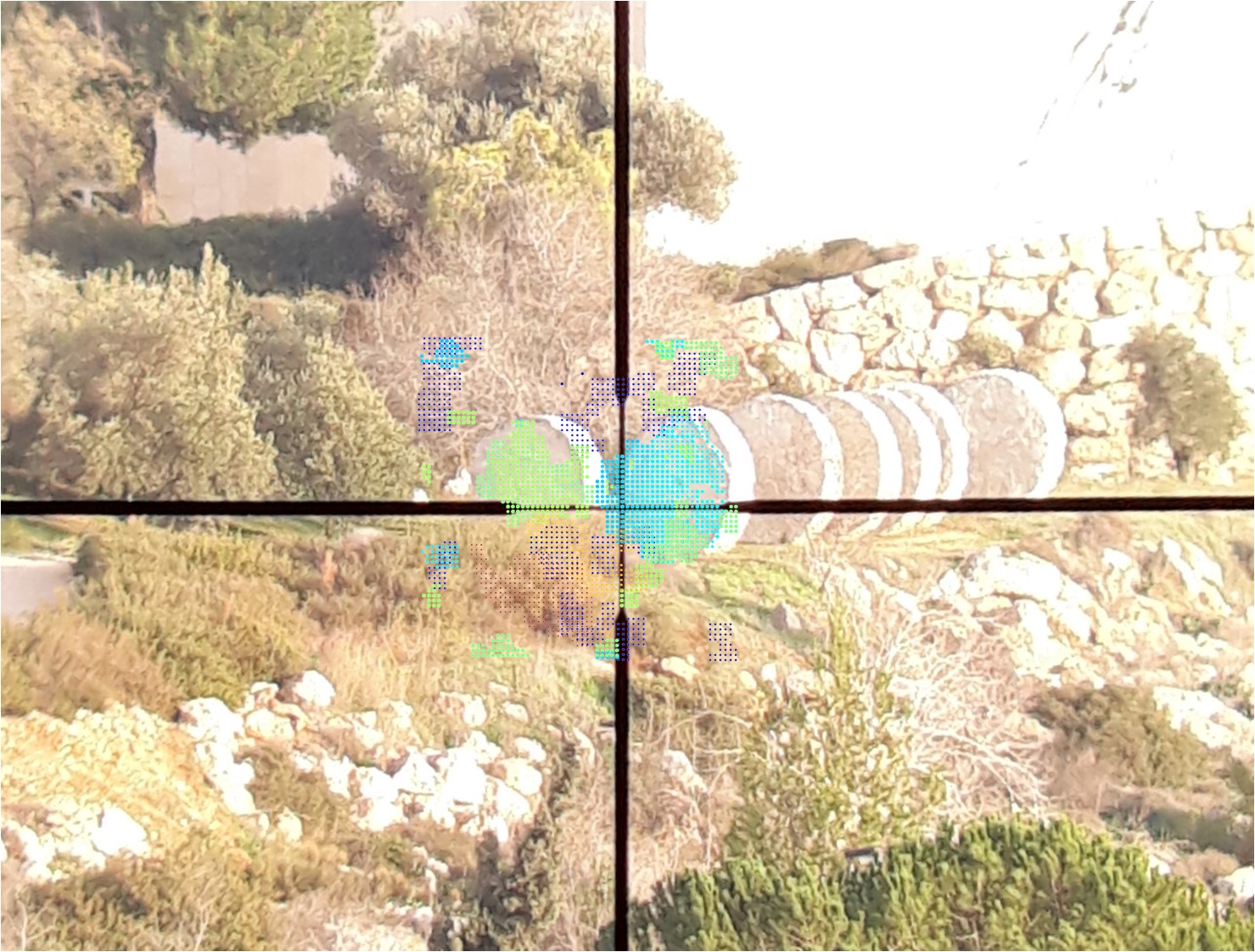}
			\label{stonesFront}}
		\hfil
		\subfloat[]{\includegraphics[height=6cm]{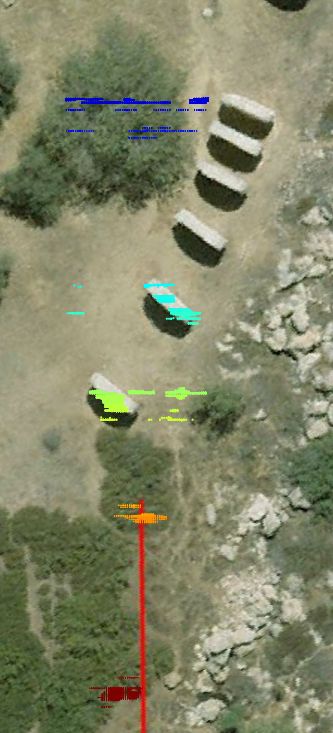}
			\label{stonesTop}}
		\hfil
		\subfloat[]{\includegraphics[height=6cm]{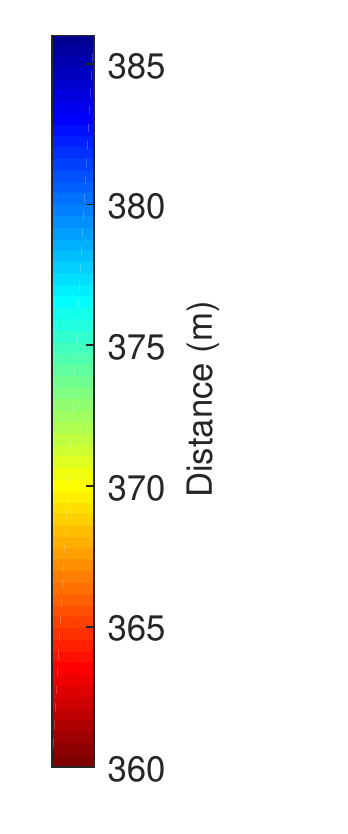}}
		\end{center}
	\caption{(\textbf{a}) Overhead image with overlay, The red line indicates the direction of view. (\textbf{b}) Picture taken through LiDAR sight with overlay. (\textbf{c}) Distance scale for overlays. }
\end{figure}

Next we aimed our LiDAR at a more ambitious target; an environmental sculpture 380 meters away. At this distance we detect less than one photons per pixel returning from the target, even at low resolution. Under these conditions a raster scan is impossible without increasing the intensity beyond eye-safe levels for visible wavelengths. In this situation our expanded beam gives a significant advantage: we can illuminate the target with much higher intensity, while leaving the energy density incident on any passerby below the strictest safety threshold. Since compressed sensing collects light from half the view field for every mask, we receive approximately 50 photons per measurement, and measuring each mask 100 times gives sufficiently accurate measurements to reconstruct an image. Figure \ref{stonesFront} shows the sculpture with an overlay of the reconstructed image and depth scale for the overlay. Direct measurement of the area captured in the image and target distance corroborates the angle of view illuminated by the diffused beam. Figure \ref{stonesTop} shows an overhead view of the sculpture, with the depth map overlay and the depth scale. 

\section{Conclusion}
We have demonstrated a LiDAR system without scanning optics, that utilizes a compressed sensing scheme, a DMD component and a photon number resolving detector. This system acquires a 3D scene with the minimal number of returning signal photons by exploiting the extreme optical linear sensitivity of the SiPM detector and the efficiency of compressed sensing approaches. Two natural scenes has been acquired, and compared to standard images and in one case also to a raster scan. The strength of the presented approach is demonstrated by the exponential improvement in the number of required masks for reconstruction, and the extremely weak detected returning signal. Further improvement is expected when better electronics and software will be implemented.

\subsection{Acknowledgments}
The authors would like to thank Yair Weiss for many fruitful discussions and ideas. 

% References
\bibliography{SPIE_LIDAR} 
\bibliographystyle{spiebib}

\end{document}